\documentclass[preprint,preprintnumbers,amsmath,amssymb]{revtex4}

\usepackage{amssymb}
\usepackage{amsmath}
\usepackage[dvips]{graphicx}

\begin{document}

\title{Optical Anisotropy of Schwarzschild Metric within Equivalent Medium Framework}

\author{Sina Khorasani and Bizhan Rashidian}
\affiliation{School of Electrical Engineering, Sharif University of Technology, P. O. Box 11365-9363, Tehran, Iran}
\email{khorasani@sina.sharif.edu}

\begin{abstract}

\noindent It is has been long known that the curved space in the presence of gravitation can be described as a non-homogeneous anisotropic medium in flat geometry with different constitutive equations. In this article, we show that the eigenpolarizations of such medium can be exactly solved, leading to a pseudo-isotropic description of curved vacuum with two refractive index eigenvalues having opposite signs, which correspond to forward and backward travel in time. We conclude that for a rotating universe, time-reversal symmetry is broken. We also demonstrate the applicability of this method to Schwarzschild metric and derive exact forms of refractive index. We derive the subtle optical anisotropy of space around a spherically symmetric, non-rotating and uncharged blackhole in the form of an elegant closed form expression, and show that the refractive index in such a pseudo-isotropic system would be a function of coordinates as well as the direction of propagation. Corrections arising from such anisotropy in the bending of light are shown and a simplified system of equations for ray-tracing in the equivalent medium of Schwarzschild metric is found.
\end{abstract}

\maketitle

\section {Introduction}

It has been shown [1,2] that the propagation of electromagnetic waves in curved space, can be described by a mathematically equivalent anisotropic medium in flat geometry. Recent uses of coordinate transformation [3] and equivalent medium theory [4] have paved the way for applications of the theory of general relativity in the domain of artificial anisotropic metamaterials. Further applications of transformation media in optical cloacking and invisibility are described in [5], and in a recent extensive review article on the geometry of light by Leonhardt and Philbin [6]. Some prior studies are also summarized in [7]. But to date and the best knowledge of the authors, no analysis of the governing electromagnetic equations for Schwarzschild metric within equivalent medium theory, has been published.

Here, the optical anisotropy of curved space is demonstrated by means of a rigorous algebraic analysis. We derive the eigenmodes of propagation and conclude that vacuum exhibits a property very much similar to pseudo-isotropic media [8], but with broken symmetry with respect to the waves travelling forward and backward in time. A simple pseudo-isotropic medium has different refractive indices along all propagation directions, but exhibits no birefringence when standard constitutive relations are used [8-10]. This broken symmetry, better known as Sagnac effect, arises from the rotation of
spacetime, and its implications in relativity and on the propagation of light has been investigated and reviewed by Schleich et. al. [11].

For this purpose, we start by inserting the constitutive relations into the Maxwell's wave equations and obtain the governing equation for eigenpolarizations. This is shown to result in a modified normal surface equation for the refractive index eigenvalue. Simplification for the pseudo-isotropic behavior gives rise to two different refractive indices with opposite signs, which are equal in magnitude for a non-rotating spacetime. We are thus led to the conclusion that, the speed of light is dependent on the local geometry of the spacetime, but at the same time, the curved space of a rotating metric is differently seen by photons, which propagate in opposite directions along the time coordinate at velocities slightly below and above \textit{c}. The difference is easily seen to be removed for non-rotating metrics. In otherwords, the time-reversal symmetry of Maxwell's equations breaks down under rotation. This conclusion is in contrast to the famous statement by Richard Feynman (1985) [12, p. 98]: ``Every particle in nature has an amplitude to move backwards in time, and therefore has an anti-particle... Photons look~exactly the same in all respects when they travel backwards in time...so they are their own anti-particles...'' It is thus here shown that Feynman's conjecture does not hold for rotating metrics.

As examples of applicability of our proposed formulation, we consider Nearly Newtonian (better known as linearized Schwarzschild metric in isotropic coordinates), Newtonian, and Schwarzschild metrics, which are all spherically symmetric and non-rotating. We derive exact closed forms for refractive index in these cases. While the Newtonian metric gives a non-homogeneous and isotropic refractive index, the latter metric is shown to result in a dependence to the radial coordinate as well as the direction of propagation and is explained by a non-homogeneous pseudo-isotropic model. We make a comparison to Einstein's results in 1911 and 1915 and show that the latter is yet subject to an important correction term.

In a review paper written in 1907 Einstein mentions that [13,14] ``\dots in the discussion that follows, we shall therefore assume the complete physical equivalence of a gravitational field and a corresponding acceleration of the reference system''. We here conclude according to the equivalence principle, that isotropy of velocity of light does not hold for non-inertial frames, which makes the space locally anisotropic, and as a result, the refractive index should depend on the direction of propagation. It is worth mentioning here that proposals of two new satellite missions, namely Gelileo Galilei (GG) [15], and Matter Wave eXplorer of Gravity (MWXG) [16], are devoted to precise interferometric tests of the equivalence principle with unprecedented accuracies.

\section{Theory}

An empty curved spacetime may be seen as an equivalent flat spacetime with a nonhomogenous and anisotropic hypothetical dielectric filled everywhere [2], the constitutive relations of which between electric \textbf{E} and magnetic \textbf{H} fields in SI units take the form [2,5]

\begin{eqnarray}
{\bf{D}} = \varepsilon _0 \left[ \varepsilon  \right]{\bf{E}} + \frac{1}{c}{\bf{w}} \times {\bf{H}}\\
{\bf{B}} = \mu _0 \left[ \mu  \right]{\bf{H}} - \frac{1}{c}{\bf{w}} \times {\bf{E}}\nonumber
\end{eqnarray}

\noindent Here, the symmetric dimensionless tensors of relative permittivity $\varepsilon$ and permeability $\mu$ are given by

\begin{equation}
\left[\varepsilon \right]=\left[\mu \right]=-\frac{\sqrt{-g} }{g_{00} } \left[g^{ij} \right]
\end{equation}

\noindent where $g^{ij}$ and $g_{ij}$ are respectively the contravariant and covariant elements of the metric tensor of space, with $g$ being the determinant of 4-metric. Also, the gyration vector \textbf{w} is defined as

\begin{equation}
{\bf w}=\frac{1}{g_{00} } \left\{g_{0i} \right\}
\end{equation}

\noindent Usually, non-rotating geometries are described with metrics having $g_{0i}=0$ for $i=1,2,3$ so that the gyration vector \textbf{w} vanishes. In fact, it has been shown that \textbf{w} is proportional to ${\bf J}\times {\bf r}$, where \textbf{J} is the angular momentum of the rotating metric [17,18].

We are now able to show that the electromagnetic curved vacuum behaves as a birefringent pseudo-isotropic medium, under the sole approximation that the variations of permittivity and permeability tensors due to the local geometry of space-time occurs on a length scale much greater than the wavelength, such that these two tensors as well as the gyration vector \textbf{w} may be taken effectively as constants. As it is shown below, time-harmonic plane-wave solutions may be sought in this case.

We start by plugging (2) into Maxwell's equations, which gives

\begin{eqnarray}
{\bf k}\times {\bf E}=\omega {\bf B}=\omega \mu _{0} \left[\mu \right]{\bf H}-k_{0} {\bf w}\times {\bf E} \\
{\bf k}\times {\bf H}=-\omega {\bf D}=-\omega \varepsilon _{0} \left[\varepsilon \right] {\bf E}-k_{0} {\bf w}\times {\bf H} \nonumber
\end{eqnarray}

\noindent Here, $\omega$ is the angular frequency, \textbf{k} is the wavevector with $k_0=\omega/c$ being the freespace wavenumber. Dividing both sides of (4) by $k_0$, we get

\begin{eqnarray}
\left(n\textbf{h}+\textbf{w}\right)\times \textbf{E}=\eta _{0} \left[\mu \right]\textbf{H} \\
\left(n\textbf{h}+\textbf{w}\right)\times \textbf{H}=-\frac{1}{\eta _{0} } \left[\varepsilon \right]\textbf{E} \nonumber
\end{eqnarray}

\noindent where $\textbf{h}=\textbf{k}/|\textbf{k}|$ is the unit vector along the propagation vector and $\eta_0$ is the intrinsic impedance of vacuum. Furthermore, $n=| \textbf{k}|/k_0$ is the index of refraction of vacuum; clearly, we have $\textbf{k}=nk_0\textbf{h}$. By defining $\textbf{q}=\textbf{h}+{\frac{1}{n}} \textbf{w}$ and $\textbf{G}=\eta _{0} \textbf{H}$, relations (5) may be rearranged as

\begin{eqnarray}
n \textbf{q}\times \textbf{E}=\left[\mu \right]\textbf{G} \\
n \textbf{q}\times \textbf{G}=-\left[\varepsilon \right]\textbf{E} \nonumber
\end{eqnarray}

\noindent Notice that $\textbf{q}$ is not a unit vector unless gyration vector vanishes, i.e. $\textbf{w}=0$. A zero gyration vector usually implies no rotation [17,18], as discussed above, however, it should be emphasized that not in all systems of coordinates a non-zero $\textbf{w}$ vector denotes true physical rotation. For instance, in the Gullstrand--Painlev\'{e} coordinates [19,20], which is connected to the Schwarzschild metric through the so-called rain transformation [21], the gyration vector does not vanish although the corresponding black hole is static and non-rotating. In fact, the time coordinate in Gullstrand--Painlev\'{e} coordinates is not the exactly the local proper time as measured by the black hole, and is obtained through a Lorentz transformation in a frame co-moving with a body being accelerated toward the black hole from rest at infinity.

The equations (6) may be combined to obtain the governing eigenpolarization equation as

\begin{eqnarray}
{\mathcal L}_{G} \textbf{G}=0 \\
{\mathcal L}_{G} =\left[\mu \right]^{-1} \textbf{q}\times \left\{\left[\varepsilon \right]^{-1} \textbf{q}\times \left(\cdot \right)\right\}+\frac{1}{n^{2} } \nonumber\\
{\mathcal L}_{E} \textbf{E}=0  \nonumber\\
{\mathcal L}_{E} =\left[\varepsilon \right]^{-1} \textbf{q}\times \left\{\left[\mu \right]^{-1} \textbf{q}\times \left(\cdot \right)\right\}+\frac{1}{n^{2} } \nonumber
\end{eqnarray}

\noindent Now, if we define the unit vector $\textbf{s}=\textbf{q}/q$, then (7) may be further simplified as

\begin{eqnarray}
{\mathcal L}\textbf{F}=0 \\
{\mathcal L}=\left[\mu \right]^{-1} \textbf{s}\times \left\{\left[\varepsilon \right]^{-1} \textbf{s}\times \left(\cdot \right)\right\}+\frac{1}{m^{2} \left(\textbf{s}\right)} \nonumber
\end{eqnarray}

\noindent Here, \textbf{F} is either of the field vectors \textbf{E} or \textbf{G}, and furthermore, the modified eigenvalue is $m\left(\textbf{s}\right)=nq\left(\textbf{s}\right)=n\left|\textbf{h}+\frac{1}{n}\textbf{w}\right|=\left|n\textbf{h}+\textbf{w}\right| =\left|\textbf{k}+k_{0} \textbf{w}\right|/k_{0} $. Notice that (8) is applicable to both of \textbf{E} or \textbf{G} because of (2).

Since the permittivity and permeability tensors are symmetric, both can be diagonalized by a proper local orthogonal transformation in 3-space, or local rotation of system of coordinates. However as it is shown later, the final refractive index may be expressed in a rotationally invariant form, meaning that such a transformation is not really necessary and does not affect the results. In such case, the normal surface equation for the modified refractive index eigenvalue \textit{m} in the local principal system of coordinates would be [8]

\begin{eqnarray}
 {\left(\frac{1}{\mu _{z} \varepsilon _{y} } -\frac{1}{m^{2} } \right)\left(\frac{1}{\mu _{y} \varepsilon _{z} } -\frac{1}{m^{2} } \right)s_{x} ^{2} +\left(\frac{1}{\mu _{x} \varepsilon _{z} } -\frac{1}{m^{2} } \right)\left(\frac{1}{\mu _{z} \varepsilon _{x} } -\frac{1}{m^{2} } \right)s_{y} ^{2} }\\
 {\quad +\left(\frac{1}{\mu _{y} \varepsilon _{x} } -\frac{1}{m^{2} } \right)\left(\frac{1}{\mu _{x} \varepsilon _{y} } -\frac{1}{m^{2} } \right)s_{z} ^{2} =\left(\frac{1}{\varepsilon _{x} } -\frac{1}{\varepsilon _{y} } \right)\left(\frac{1}{\mu _{x} } -\frac{1}{\mu _{y} } \right)\frac{s_{x} ^{2} s_{y} ^{2} }{\varepsilon _{z} \mu _{z} } } \nonumber \\
 {\quad +\left(\frac{1}{\varepsilon _{y} } -\frac{1}{\varepsilon _{z} } \right)\left(\frac{1}{\mu _{y} } -\frac{1}{\mu _{z} } \right)\frac{s_{y} ^{2} s_{z} ^{2} }{\varepsilon _{x} \mu _{x} } +\left(\frac{1}{\varepsilon _{z} } -\frac{1}{\varepsilon _{x} } \right)\left(\frac{1}{\mu _{z} } -\frac{1}{\mu _{x} } \right)\frac{s_{z} ^{2} s_{x} ^{2} }{\varepsilon _{y} \mu _{y} } }\nonumber
\end{eqnarray}

\noindent The equation (9) is in general a biquadratic equation for the eigenvalue $m$, and in general has two distinct roots for $m^2$ along all directions of $\textbf{s}$, but the so-called optical axes. We have discussed the conditions under which (9) leads to exactly two optical axes corresponding to a biaxial medium, or one optical axis corresponding to a uniaxial medium. However, when the condition

\begin{equation}
\frac{\mu _{x} }{\varepsilon _{x} } =\frac{\mu _{y} }{\varepsilon _{y} } =\frac{\mu _{z} }{\varepsilon _{z} }
\end{equation}

\noindent holds, which is easily inferred from (2) for the curved empty space, then (9) allows double roots along every possible direction $\textbf{s}$. This situation is referred to as pseudo-isotropic medium, in which the medium has no birefringence but is still anisotropic. For a pseudo-isotropic medium, the space looks identical along perpendicular directions everywhere, yet different along various propagation directions. Hence, all eigenpolarizations are degenerate along all propagation directions, and every polarization satisfying $\textbf{k}\cdot \textbf{B}=\textbf{k}\cdot \textbf{D}=\textbf{B}\cdot \textbf{E}=\textbf{D}\cdot \textbf{H}=0$ together with the constitutive relations, would be an eigenpolarization [8-10].

After considerable but straightforward algebra, it is possible to show that the solutions of (9) for a pseudo-isotropic medium are given by

\begin{equation}
m^{2} =\frac{\varepsilon _{x} \mu _{y} \mu _{z} }{\mu _{x} s_{x} ^{2} +\mu _{y} s_{y} ^{2} +\mu _{y} s_{y} ^{2} } =\frac{\varepsilon _{x} \mu _{y} \mu _{z} }{\textbf{s}\cdot \left[\mu \right]\cdot \textbf{s}}
\end{equation}

\noindent Notice that cyclic permutations of \textit{x}, \textit{y}, and \textit{z} indices in (11) give rise to the same result when (10) holds. For the case of our interest, where $\left[\varepsilon \right]=\left[\mu \right]$, (11) may be rewritten in the rotationally invariant form

\begin{equation}
m^{2} =\frac{\left|\mu \right|}{\textbf{s}\cdot \left[\mu \right]\cdot \textbf{s}} =\frac{\left|\varepsilon \right|}{\textbf{s}\cdot \left[\varepsilon \right]\cdot \textbf{s}}
\end{equation}

\noindent where $\left|\cdot \right|$ represents the determinant operation. In the absence of rotation \textbf{w}=0 as well as curved geometry, we have \textit{m}=\textit{n} and $\left[\varepsilon \right]=\left[\mu \right]=1$. Then the plausible refractive indices from (12) are given by $n=\pm 1$. Non-trivially, the negative solution corresponds to the waves travelling backward in time (Appendix A); some earlier works had postulated negative refraction [22], which was later shown to be wrong [23]. This situation states in other words that in the absence of gravitational field, the electromagnetic field looks the same for photons and anti-photons. In contrast to positrons, anti-photons may be regarded as photons traveling backward in time [12]. As we shall see below, the symmetry is broken in the presence of rotating gravitational field, and as a result, photons and anti-photons become different at least with regard to their velocities.

The actual solutions of (12) for the eigenvalue \textit{n} become complicated when we take the dependency of \textit{m} on \textbf{s} into account. By definition, $m^{2} \left(\textbf{s}\right)=\left|n\textbf{h}+\textbf{w}\right|^{2} $ and $\textbf{s}=\textbf{q}/q$. Then, we have

\begin{equation}
\left|n\textbf{h}+\textbf{w}\right|^{2} =\frac{\left|\mu \right|}{\left(\frac{\textbf{q}}{q} \right)\cdot \left[\mu \right]\cdot \left(\frac{\textbf{q}}{q} \right)}
\end{equation}

\noindent which by noting $\textbf{q}=\textbf{h}+{\frac{1}{n}} \textbf{w}$ can be further simplified to obtain the master eigenvalue equation for the refractive index of curved vacuum as

\begin{equation}
\left(n\textbf{h}+\textbf{w}\right)\cdot \left[\xi \right]\cdot \left(n\textbf{h}+\textbf{w}\right)=1
\end{equation}

\noindent Here, $\left[\xi \right]=\left[\mu \right]/\left|\mu \right|$. Expanding (14) in its components gives

\begin{equation}
\xi _{x} \left(nh_{x} +w_{x} \right)^{2} +\xi _{y} \left(nh_{y} +w_{y} \right)^{2} +\xi _{z} \left(nh_{z} +w_{z} \right)^{2} =1
\end{equation}

\noindent which can be rearranged as

\begin{eqnarray}
An^{2} +2Bn-C=0\\
A=\textbf{h}\cdot \left[\xi \right]\cdot \textbf{h} \nonumber \\
B=\textbf{h}\cdot \left[\xi \right]\cdot \textbf{w} \nonumber \\
C=1-\textbf{w}\cdot \left[\xi \right]\cdot \textbf{w} \nonumber
\end{eqnarray}

\noindent Therefore, the exact solutions are

\begin{eqnarray}
n_{1} =+\frac{\sqrt{AC+B^{2} } -B}{A} =+\frac{\sqrt{\left(\textbf{h}\cdot \left[\xi \right]\cdot \textbf{h}\right)\left(1-\textbf{w}\cdot \left[\xi \right]\cdot \textbf{w}\right)+\left(\textbf{h}\cdot \left[\xi \right]\cdot \textbf{w}\right)^{2} } -\textbf{h}\cdot \left[\xi \right]\cdot \textbf{w}}{\textbf{h}\cdot \left[\xi \right]\cdot \textbf{h}} \\
n_{2} =-\frac{\sqrt{AC+B^{2} } +B}{A} =-\frac{\sqrt{\left(\textbf{h}\cdot \left[\xi \right]\cdot \textbf{h}\right)\left(1-\textbf{w}\cdot \left[\xi \right]\cdot \textbf{w}\right)+\left(\textbf{h}\cdot \left[\xi \right]\cdot \textbf{w}\right)^{2} } +\textbf{h}\cdot \left[\xi \right]\cdot \textbf{w}}{\textbf{h}\cdot \left[\xi \right]\cdot \textbf{h}} \nonumber
\end{eqnarray}

If deviation in metric from the flat Minkowskian geometry is not too strong, then the terms being second order in \textbf{w} can be dropped and (17) can be approximated as

\begin{eqnarray}
n_{1} \approx +\left(\textbf{h}\cdot \left[\xi \right]\cdot \textbf{h}\right)^{-{\frac{1}{2}} } -\frac{\textbf{h}\cdot \left[\xi \right]\cdot \textbf{w}}{\textbf{h}\cdot \left[\xi \right]\cdot \textbf{h}} \\
n_{2} \approx -\left(\textbf{h}\cdot \left[\xi \right]\cdot \textbf{h}\right)^{-{\frac{1}{2}} } -\frac{\textbf{h}\cdot \left[\xi \right]\cdot \textbf{w}}{\textbf{h}\cdot \left[\xi \right]\cdot \textbf{h}} \nonumber
 \end{eqnarray}

\noindent Note that (17) take on fairly simple forms when \textbf{w}=0, and thus we get the \textit{exact} expressions

\begin{eqnarray}
n_{1} =+\left(\textbf{h}\cdot \left[\xi \right]\cdot \textbf{h}\right)^{-{\frac{1}{2}} }  \\
n_{2} =-\left(\textbf{h}\cdot \left[\xi \right]\cdot \textbf{h}\right)^{-{\frac{1}{2}} } \nonumber
\end{eqnarray}

\noindent This shows that for non-rotating metrics when described in proper coordinate system, time-reversal symmetry holds. It should be also added that a prior study [24] has shown that vanishing birefringence is necessary for consistency, where by birefringence the authors of [24] mean the existence of two refractive indices for forward-propagation in time, corresponding to two light cones. Our study also justifies this result, and furthermore we notice that two inequal roots are given by (17), although they may differ only within a sign under some circumstances. Hence, there is one light cone for forward propagation and another light cone for backward propagation in time.

Time-reversal of Maxwell's equations plus two orthogonal polarizations allow for four distinct refractive index eigenvalues. Notice that for standard anisotropic media, from the solution of normal-surface equation we normally have two pairs of eigenvalues, $n_{1} =-n_{3} $ and $n_{2} =-n_{4} $ with either $n_{1} \ge n_{2} >0$ or $n_{2} \ge n_{1} >0$, where the equality sign holds only along optical axes; it is clear that the time-reversal symmetry holds. This is while in the pseudo-isotropic medium of our interest, we have two double roots given by $n_{1} =n_{3} $ and $n_{2} =n_{4} $ with either $n_{2} \le -1<0<n_{1} \le 1$, or $-1\le n_{2} <0<1\le n_{1} $ depending on the sense of rotation; hence, in general, the time-reversal symmetry does not hold. This is while for non-rotating metrics we have the further simplification $n_{1} =-n_{2} $, corresponding to vanishing birefringence and preserving time-reversal symmetry. Based on the definition, a pseudoisotropic medium has no birefringence at all, but the refractive index can still be dependent on the direction of propagation.

When $\textbf{w}\ne 0$, we may still note that \textbf{h} is a unit vector, and thus (18) can be still simplified further in the weak gravitational field limit as

\begin{eqnarray}
n_{1} \approx +1+{\frac{1}{2}} \Delta n\\
n_{2} \approx -1+{\frac{1}{2}} \Delta n\nonumber
\end{eqnarray}

\noindent corresponding respectively to photons and anti-photons; here, we define $\Delta n=-2\textbf{h}\cdot \left[\xi \right]\cdot \textbf{w}$. Firstly, it can be seen that the curved vacuum exhibits a local time-reversal asymmetry given by $\left|\Delta n\right|=\left|n_{1} -n_{2} \right|$, which is roughly a linear function of gravitational potential (as shown below). Secondly, photons travel at a speed slightly below (above) the speed of light in flat vacuum \textit{c}, while anti-photons travel at a speed slightly above (below) \textit{c}, if the direction of propagation is anti-parallel (along) to the rotation of the universe, where $\Delta n>0$ ($\Delta n<0$).

Now as it is shown below for Nearly Newtonian, Newtonian, and Schwarzschild metrics, the equations (19) are compared to the predictions made by Einstein for weak gravitational fields [25-27].

\section{Examples}

\noindent In this section we consider two important cases, which both correspond to spherically symmetric non-rotating universes: Newtonian and Schwarzschild metrics, where two versions of Newtonian metric are discussed. Rotation can be exactly implemented through (17), however, the resulting expressions are too complicated and hence are not discussed for the sake of convenience.

\subsection{Nearly Newtonian and Newtonian Metrics}

The spherically symmetric Nearly Newtonian metric [28, p. 445], or the so-called linearized Schwarzschild metric in isotropic coordinates is given by the line element

\begin{equation}
ds^{2} =-c^{2} \left(1-\frac{r_{s} }{r} \right)dt^{2} +\left(1+\frac{r_{s} }{r} \right)\left(dx^{2} +dy^{2} +dz^{2} \right)
\end{equation}

\noindent where $r_{s} =2GM/c^{2} $ is referred to as the Schwarzschild radius of the star with \textit{M} and \textit{G} respectively being its mass and gravitational constant, and $dl^{2} =dx^{2} +dy^{2} +dz^{2} $ is the spacelike path element. This metric is an approximate solution of Einstein field equations, but to high precision most stars are static and spherical [28, p. 446], so that (21) is applicable. Landau and Lifshitz [29] report the Newtonian metric in the slightly different form

\begin{equation}
ds^{2} =-c^{2} \left(1-2\frac{r_{s} }{r} \right)dt^{2} +\left(dx^{2} +dy^{2} +dz^{2} \right)
\end{equation}

\noindent The solution (19) may be used to all three above cases and we readily obtain

\begin{eqnarray}
n=\pm \frac{\left(r+r_{s} \right)^{{\tfrac{5}{2}} } }{r^{{\tfrac{3}{2}} } \left(r-r_{s} \right)} =\pm \left(\frac{r}{r_{s} } \right)^{-{\tfrac{3}{2}} } \left(\frac{r}{r_{s} } -1\right)^{-1} \left(\frac{r}{r_{s} } +1\right)^{{\tfrac{5}{2}} } \\
n=\pm \frac{r^{{\tfrac{1}{2}} } }{\left(r-2r_{s} \right)^{{\tfrac{1}{2}} } } =\pm \left(1-2\frac{r}{r_{s} } \right)^{-{\tfrac{1}{2}} } \nonumber
\end{eqnarray}

\noindent respectively for the metrics (21) and (22). Denoting $\Phi =-r_{s} /r$, we get

\begin{eqnarray}
n=\pm \frac{\left(1-\Phi \right)^{{\frac{5}{2}} } }{1+\Phi } \\
n=\pm \frac{1}{\left(1+2\Phi \right)^{{\frac{1}{2}} } } \nonumber
\end{eqnarray}

\noindent These expression blow up to infinity for $r\to r_{s} $ as $\left(r-r_{s} \right)^{-1} $, but approach in magnitude to unity as $r\to \infty $. In the limit of small normalized gravitational potential $r>>r_{s} $, however, we get

\begin{eqnarray}
\left|n\right|\approx 1-{\frac{7}{2}} \Phi \\
\left|n\right|\approx 1-\Phi \nonumber
\end{eqnarray}

\noindent The latter result agrees to that of Einstein's 1911 early prediction [25]. A few years later, however, he showed that the change in refractive index should have been given by $\left|n\right|\approx 1-2\Phi $ [26,27]. Now, it is discussed below that his correction factor of 2 was still inaccurate.

\subsection{Schwarzschild Metric}

Schwarzschild metric is known by Birkhoff's 1932 theorem [28, p.843], to be the most general solution of Einstein field equations under spherical symmetry and no rotation. Schwarzschild metric has been written in many coordinate systems which are all connected through transformation, including isotropic, synchronous (Lemaitre-Rylov's), Eddington-Finkelstein, Kruskal-Szekeres', harmonic, as well as Gullstrand--Painlev\'{e} coordinates [19,20,30].  In its basic form, it is given by [28, p. 607]

\begin{equation}
ds^{2} =-c^{2} \left(1-\frac{r_{s} }{r} \right)dt^{2} +\frac{dr^{2} }{1-\frac{r_{s} }{r} } +r^{2} \left(d\theta ^{2} +\sin ^{2} \theta d\phi ^{2} \right)
\end{equation}

\noindent where $\left(r,\theta ,\phi \right)$ constitute the standard spherical polar coordinates. Recently, Virbharda has revisited the problem of light deflection in Schwarzschild geometry [31-33].

The difficulty in working with this metric arises from the fact that the spacelike path element $dl^{2} =dr^{2} +r^{2} d\Omega ^{2} $ where $d\Omega ^{2} =d\theta ^{2} +\sin ^{2} \theta d\phi ^{2} $, does not appear explicitly in the metric. To overcome this difficulty, the common practice is to express (26) in a similar form to (21) using the so-called isotropic coordinates [28, p. 840], through the transformation

\begin{equation}
r=\rho \left(1+\frac{r_{s} }{2\rho } \right)^{2}
\end{equation}

\noindent where $\rho$ is called the isotropic radial coordinate. This allows to rewrite the metric as [34]

\begin{equation}
ds^{2} =-c^{2} \frac{\left(1-\frac{r_{s} }{2\rho } \right)^{2} }{\left(1+\frac{r_{s} }{2\rho } \right)^{2} } dt^{2} +\left(1+\frac{r_{s} }{2\rho } \right)^{4} \left[d\rho ^{2} +\rho ^{2} \left(d\theta ^{2} +\sin ^{2} \theta d\phi ^{2} \right)\right]
\end{equation}

\noindent which allows one to readily obtain [34,35]

\begin{equation}
n\left(\rho \right)=\left(1-\frac{r_{s} }{2\rho } \right)^{-1} \left(1+\frac{r_{s} }{2\rho } \right)^{3}
\end{equation}

\noindent This result can be approximated in the limit of large $\rho$ as

\begin{equation}
n\left(\rho \right)\approx 1+2\frac{r_{s} }{\rho } \approx 1+2\frac{r_{s} }{r} ,\quad \rho >>r_{s}
\end{equation}

\noindent By comparing (30) with (24) the Einstein's correction factor of 2 becomes evident. There is no such apparent anisotropy in (30), as obtained from transformation to isotropic coordinates.

\noindent But the new radial coordinate $\rho$ in isotropic coordinates has no direct physical meaning, while the radial coordinate in (26) is actually the circumferential radius [36,37] (proper circumference divided by $2\pi$). Therefore, we stick to the Schwarzschild coordinates to avoid wrong conclusions. However the conversion process to quasi-Minkowskian coordinates needs some care, and construction of the proper metric needs the following transformations [36, p. 181]. First, the metric (26) is rewritten as

\begin{equation}
ds^{2} =-c^{2} \left(1-\frac{r_{s} }{r} \right)dt^{2} +\frac{r_{s} }{r-r_{s} } dr^{2} +dl^{2}
\end{equation}

\noindent and then we note that $r=\sqrt{x^{2} +y^{2} +z^{2} } $. This enables us to write down

\begin{equation}
dr^{2} =\frac{x_{i} x_{j} }{r^{2} } dx^{i} dx^{j}
\end{equation}

\noindent in which \textit{i},\textit{j}=1,2,3 denote \textit{x}, \textit{y}, and \textit{z} coordinates. Hence (31) can be rewritten as

\begin{equation}
ds^{2} =-c^{2} \left(1-\frac{r_{s} }{r} \right)dt^{2} +\left(\frac{r_{s} }{r-r_{s} } \frac{x_{i} x_{j} }{r^{2} } +\delta _{ij} \right)dx^{i} dx^{j}
\end{equation}

\noindent This metric is obviously non-diagonal, but can be diagonalized everwhere through orthogonal transformations in 3-space. Since eigenvalues of the resulting tensors are degenerate, Gram-Schmidt ortho-normalization was applied to find the principal axes of coordinates. It should be noted that the propagation vector $\textbf{h}$ should be rotated accordingly to obtain correct results. Furthermore, (19) are exact as $\textbf{w}=0$.

Since $\textbf{h}$ is a unit vector it can be described by the spherical polar angles $\left(\psi ,\chi \right)$ in the original non-rotated reference frame. A tedious but straightforward calculation then gives the eigenvalues of propagation as

\begin{equation}
\left|n\left(\textbf{r}\right)\right|=\frac{\left(\frac{r}{r_{s} } \right)^{2} }{\left(\frac{r}{r_{s} } -1\right)^{{\tfrac{3}{2}} } \sqrt{\frac{r}{r_{s} } -1+\left[\cos \left(\phi -\chi \right)\cos \left(\theta -\psi \right)+2\sin ^{2} \left(\frac{\chi -\phi }{2} \right)\cos \theta \cos \psi \right]^{2} } }
\end{equation}

\noindent Spherical symmetry makes the absolute choice of the angles $\left(\psi ,\chi \right)$ irrelevant, in the sense that we may set the \textit{z}-axis along the position coordinate \textbf{r} and choose $\phi =\theta =0$. Hence, we get the \textit{exact} yet fairly simple expression

\begin{equation}
\left|n\left(\textbf{r}\right)\right|=\frac{\left(\frac{r}{r_{s} } \right)^{2} }{\left(\frac{r}{r_{s} } -1\right)^{{\tfrac{3}{2}} } \sqrt{\frac{r}{r_{s} } -\sin ^{2} \psi } }
\end{equation}

\noindent where the azimuthal angle of propagation $\psi$ is measured with respect to the position vector \textbf{r}. Again for $r\to r_{s}$ the refractive index blows up respectively as $\left(r-r_{s} \right)^{-{\tfrac{3}{2}} } $ and $\left(r-r_{s} \right)^{-2} $, if $\psi \ne \pm {\tfrac{\pi }{2}} $ and $\psi =\pm {\frac{\pi }{2}} $.  However, it approaches in magnitude to unity as $r\to \infty $, since the metric relaxes to that of the Minkowskian in the limit of infinite radius.

Finally, we adopt $\Phi =-r_{s} /r$, and rewrite (35) as

\begin{equation}
\left|n\left(\textbf{r}\right)\right|=\left(1+\Phi \right)^{-{\tfrac{3}{2}} } \left(1+\Phi \sin ^{2} \psi \right)^{-{\tfrac{1}{2}} }
\end{equation}

\noindent This result can be rewritten in the limit of small $\Phi $ (weak gravitational field) as

\begin{equation}
\left|n\left(\textbf{r}\right)\right|\approx 1-\frac{3+\sin ^{2} \psi }{2} \Phi
\end{equation}

\noindent Since $\psi$ is actually the angle made by the position vector $\textbf{r}$ and wavevector $\textbf{k}$, the above equation
can be also put in the more convenient form

\begin{equation}
\left|n\left(\textbf{r},\textbf{k}\right)\right|\approx 1-\left[2 -\frac{(\textbf{k}\cdot \textbf{r})^2}{2k^2r^2}\right] \Phi
=1-(2-\frac{h_{r}^2}{2}) \Phi
\end{equation}

\noindent where $h_r=k_r/k=\textbf{h}\cdot\textbf{r}/r$. The correction factor to the Einstein's 1915 result as shown in (30), hence actually varies between 3/2 and 2 depending on the angle of propagation. The light ray evidently keeps passing on the plane defined by the vectors $\textbf{r}$ and $\textbf{k}$. Then $h_r=0$ or $\sin ^{2} \psi =1$ holds only at the nearest point to the center of the star in the light trajectory, where (30) is accurate, while at farther points away from the center of the star, we approach $h_r=1$ or $\sin ^{2} \psi =0$.

Another conclusion is that this anisotropy is expected to be present everywhere around a massive object, so that the change in refractive index by changing the direction of propagation from $\psi ={\tfrac{\pi }{2}} $ to $\psi =0$, could reach as high as $\left|\Phi \right|/2=\left(GM/c^{2} \right)/r$. Based on the estimates given in [25, p. 459], this figure should be of the order of $10^{-8} $ for an experiment done at Earth's distance from Sun, while it would be only about $6\times 10^{-10} $ at the surface of Earth when the gravity of Sun is neglected. Finally, the equivalence principle states that gravity should be indistinguishable from pure acceleration. Therefore, accelerated (non-inertial) frames should have optical anisotropy, meaning that the speed of light along various directions is not necessarily the same for a non-inertial observer.

\subsubsection{Deflection of Light}

Here, we investigate the effect of anisotropy in deflection of light around a massive Schwarzschild object. When the coordinates are
expressed in the isotropic system of coordinates, the angle of deflection is simply given by [38-40]

\begin{equation}
\alpha=\frac{2r_s}{r_0}
\end{equation}

\noindent where $r_0$ is the closest distance of approach of the light ray to the star. However, when the effects of anisotropy are included,
one must solve the associated equations of motion in the Schwarzschild metric. A classic solution to this problem in Schwarzschild geometry has been proposed by Weinberg [36], and more recently been revisited by Virbhadra [31-33]. This results in the expression

\begin{equation}
\alpha=\frac{2r_s}{r_0}+\frac{r_s^2}{r_0^2}(\frac{15\pi}{16}-1)+\dots
\end{equation}

\noindent where by comparing to (39), it becomes possible that the first correction term may be attributed to the anisotropy of Schwarzschild metric.

\subsubsection{Ray Tracing Equations}

Propagation of light rays within the equivalent medium framework can be approximated by geometrical optics. This results in two coupled sets of equations, together describing the equations of motion for the beam [4,41,42]. The Quasi-Isotropic Approximation (QIA) [43,44] for weakly anisotropic media is best suited to deal with this problem when the rotation of polarization is sought. This is particularly appealing, since the equivalent medium of vacuum is both pseudo-isotropic with vanishing birefringence and weakly anisotropic.  Anyway, the exact equations of motion take the form

\begin{eqnarray}
\frac{d\textbf{r}}{d\tau}=+\frac{\partial H}{\partial \textbf{k}}\\
\frac{d\textbf{k}}{d\tau}=-\frac{\partial H}{\partial \textbf{r}} \nonumber
\end{eqnarray}

\noindent Here, the Hamiltonian $H$ is expressed by $H=f(\textbf{r})(\textbf{k}\cdot\left[\xi\right]\cdot\textbf{k}-\left|\xi\right|)$, with $f(\textbf{r})$ being an arbitrary function of position. The above set of equations result in six nonlinearly coupled first-order differential equations, which normally admit only numerical solution. A great deal of simplification is possible for the stationary case when the free parameter $\tau=l$ is the same as arc length covered by the ray [37]

\begin{eqnarray}
\frac{d\textbf{r}}{dl}=\frac{\textbf{b}(\textbf{r},\textbf{k})}{\left|\textbf{b}(\textbf{r},\textbf{k}) \right|}\\
\frac{d\textbf{k}}{dl}=-\frac{\left|\textbf{k}\right|}{\left|\textbf{b}(\textbf{r},\textbf{k}) \right|}\frac{\partial n^{-1}(\textbf{r},\textbf{k})}{\partial \textbf{r}}\nonumber
\end{eqnarray}

\noindent where

\begin{eqnarray}
\textbf{b}(\textbf{r},\textbf{k}) = \frac{\partial \left|\textbf{k}\right|n^{-1}(\textbf{r},\textbf{k})}{\partial\textbf{k}}
\end{eqnarray}

\noindent Now from (38), we may insert $n^{-1}(\textbf{r},\textbf{k})\approx 1+(2-\frac{1}{2}h_{r}^2) \Phi$, which by defining the unit radial vector $\hat{r}=\textbf{r}/r$ results in $\textbf{b}(\textbf{r},\textbf{k}) = \textbf{h}\left[1+\Phi\left(2+\frac{1}{2}h_r^2\right)\right]-h_r\Phi \hat{r}$. We also have $b(\textbf{r},\textbf{k}) = \sqrt{\textbf{b}(\textbf{r},\textbf{k})\cdot\textbf{b}(\textbf{r},\textbf{k})}$, which after removing second- and higher-order terms in $\Phi$ and applying binomial expansion takes the form $b(\textbf{r},\textbf{k}) \approx 1+ \Phi (2-\frac{1}{2}h_r^2)$. Furthermore, the gradient in the right-hand-side of (42) simplifies as

\begin{equation}
\frac{\partial n^{-1}(\textbf{r},\textbf{k})}{\partial \textbf{r}}\approx -\frac{\Phi}{r}\left[\left(2-\frac{3}{2}h_r^2\right)\hat{r}+h_r \textbf{h}\right]
\end{equation}

\noindent Therefore, we obtain the ray tracing equations for the equivalent medium of Schwarzschild geometry, being correct to $O(\Phi)$ as

\begin{eqnarray}
\frac{d\textbf{r}}{dl}\approx \left(1+h_r^2\Phi\right)\textbf{h}-h_r\Phi\hat{r} \\
\frac{d\textbf{k}}{dl}=\frac{\Phi}{r}\left|\textbf{k}\right|\left[\left(2-\frac{3}{2}h_r^2\right)\hat{r}+h_r \textbf{h}\right]\nonumber
\end{eqnarray}

\section{Conclusions}

\noindent In summary, we have discussed the optical anisotropy of vacuum subject to a gravitational field. Exact expressions for the refractive indices are obtained, and shown to correspond to waves traveling forward and backward in time. These two eigenvalues are different in magnitude for a rotating spacetime, meaning that photons and anti-photons look different in rotating curved spaces, travelling at velocities slightly below and above \textit{c}. We also derived exact and closed-form expressions of the refractive index for Nearly Newtonian, Newtonian, and Schwarzschild metrics, and showed that the latter metric gives rise to a locally pseudo-isotropic universe. It has been furthermore proved that a strong local anisotropy is predicted to exist in the equivalent medium description of curved geometry, which shows that the Einstein's 1915 correction factor might have still been inaccurate. Based on the equivalence principle, gravity is indistinguishable from pure acceleration. Hence, accelerated non-inertial frames induce optical anisotropy in the speed of light.

\begin{acknowledgments}
One of the authors (S. Khorasani) appreciates discussions of this work with Professor Bahram Mashhoon at the University of Missouri. The authors thank the anonymous reviewers of this article for providing constructive and valuable comments. This paper is dedicated in loving memory of the inspiring teacher of the School of Electrical Engineering at Sharif University of Technology, the late Professor Dr. Kasra Barkeshli (1961-2005) [45].
\end{acknowledgments}

\appendix
\section{Negative Refractive Index Does not Imply Negative Refraction}

We here show that existence of a negative refractive index does not imply negative refraction. This fact is more than trivial, but can be easily observed by checking the Maxwell's equations for simple classical vacuum with flat geometry:

\begin{eqnarray}
{\nabla \times \textbf{E}=-\frac{\partial }{\partial t} \textbf{B}=-\mu _{0} \frac{\partial }{\partial t} \textbf{H}} \\
{\nabla \times \textbf{H}=+\frac{\partial }{\partial t} \textbf{D}=+\varepsilon _{0} \frac{\partial }{\partial t} \textbf{E}} \nonumber
\end{eqnarray}

\noindent Substitution of solutions varying as $\exp \left[j\left(\omega t-\textbf{k}\cdot \textbf{r}\right)\right]$ gives

\begin{eqnarray}
{\textbf{k}\times \textbf{E}=+\mu _{0} \omega \textbf{H}} \nonumber \\
{\textbf{k}\times \textbf{H}=-\varepsilon _{0} \omega \textbf{E}} \nonumber
\end{eqnarray}

\noindent The wave equation for fields may thus be written as

\begin{eqnarray}
{\textbf{k}\times \left(\textbf{k}\times \textbf{E}\right)=-\omega ^{2} \mu _{0} \varepsilon _{0} \textbf{E}} \nonumber\\
{\textbf{k}\times \left(\textbf{k}\times \textbf{H}\right)=-\omega ^{2} \mu _{0} \varepsilon _{0} \textbf{H}} \nonumber
\end{eqnarray}

\noindent Noting that from Maxwell's divergence equations we have $\textbf{k}\cdot \textbf{E}=\textbf{k}\cdot \textbf{H}=0$, the above may be rewritten as

\begin{eqnarray}
{\left(\textbf{k}\cdot \textbf{k}\right)\textbf{E}=\omega ^{2} \mu _{0} \varepsilon _{0} \textbf{E}=\frac{\omega ^{2} }{c^{2} } \textbf{E}} \nonumber \\
{\left(\textbf{k}\cdot \textbf{k}\right)\textbf{H}=\omega ^{2} \mu _{0} \varepsilon _{0} \textbf{H}=\frac{\omega ^{2} }{c^{2} } \textbf{H}} \nonumber
\end{eqnarray}

\noindent Both equations are satisfied if the following dispersion equation holds

\begin{equation}
\textbf{k}\cdot \textbf{k}=k^{2} =\frac{\omega ^{2} }{c^{2} }
\end{equation}

\noindent For a fixed direction of \textbf{k}, the \textit{two} solutions to the above \textit{second-order} equation are $\textbf{k}_{1} =n_{1} \frac{\left|\omega \right|}{c} \hat{s}$ and $\textbf{k}_{2} =n_{2} \frac{\left|\omega \right|}{c} \hat{s}$, where $\hat{s}$ is a unit vector, $n_{1} ^{2} =n_{2} ^{2} =1$, and $\left|\cdot \right|$ denotes the absolute value. Clearly, we have the acceptable solutions as $\textbf{k}_{1} =+\frac{\omega }{c} \hat{s}$ and $\textbf{k}_{2} =-\frac{\omega }{c} \hat{s}$. Because of the obtained forms, we may choose the sign of $\omega $ arbitrarily as positive, but will keep it fixed throughout. We can show that both solutions do have physical interpretation.

Note that both solutions to the dispersion equation (A2) are acceptable while there is evidently no negative refraction. Now we insert the solutions in the original Maxwell's equations to get

\begin{eqnarray}
{+\frac{\omega }{c} \hat{s}\times \textbf{E}=+\mu _{0} \omega \textbf{H}\quad \Rightarrow \quad \hat{s}\times \textbf{E}=+\textbf{G}} \\
{+\frac{\omega }{c} \hat{s}\times \textbf{H}=-\varepsilon _{0} \omega \textbf{E}\quad \Rightarrow \quad \hat{s}\times \textbf{G}=-\textbf{E}} \nonumber
\end{eqnarray}

\begin{eqnarray}
{-\frac{\omega }{c} \hat{s}\times \textbf{E}=+\mu _{0} \omega \textbf{H}\quad \Rightarrow \quad \hat{s}\times \textbf{E}=-\textbf{G}} \\
{-\frac{\omega }{c} \hat{s}\times \textbf{H}=-\varepsilon _{0} \omega \textbf{E}\quad \Rightarrow \quad \hat{s}\times \textbf{G}=+\textbf{E}} \nonumber
\end{eqnarray}

\noindent where $\eta _{0} =c\mu _{0} $ and $\textbf{G}=\eta _{0} \textbf{H}$.

Notice that the first set of equations (A3) corresponds to a right-handed coordinate system constructed of $\left(\hat{s},\hat{e},\hat{g}\right)$, while (A4) corresponds to a left-handed coordinate system $\left(\hat{s},\hat{e},\hat{g}\right)$, in which $\hat{e}=\textbf{E}/\left|\textbf{E}\right|$ and $\hat{g}=\textbf{G}/\left|\textbf{G}\right|=\textbf{H}/\left|\textbf{H}\right|$ are field unit vectors. This is while the left-handed solution (A4) could be obtained equivalently from the ansatz $\exp \left[j\left(\omega t+\textbf{k}\cdot \textbf{r}\right)\right]$, too.

In summary, (A1) admits two different particular solutions, given by

\begin{eqnarray}
{\textbf{E}\left(\textbf{r},t\right)=Re\left[\textbf{E}_{0} \exp \left[j\left(\omega t-\textbf{k}\cdot \textbf{r}\right)\right]\right]=\left|\textbf{E}_{0} \right|\cos \left[\omega t-\textbf{k}\cdot \textbf{r}+\measuredangle E_{0} \right]} \nonumber \\
{\textbf{H}\left(\textbf{r},t\right)=Re\left[\textbf{H}_{0} \exp \left[j\left(\omega t-\textbf{k}\cdot \textbf{r}\right)\right]\right]=\left|\textbf{H}_{0} \right|\cos \left[\omega t-\textbf{k}\cdot \textbf{r}+\measuredangle H_{0} \right]} \nonumber
\end{eqnarray}

\noindent with either $\textbf{H}_{0} =+\hat{s}\times \textbf{E}_{0} /\eta _{0} =\textbf{k}\times \textbf{E}_{0} /\omega \mu _{0} $ and $\textbf{k}=+\omega \hat{s}/c$, or $\textbf{H}_{0} =-\hat{s}\times \textbf{E}_{0} /\eta _{0} =\textbf{k}\times \textbf{E}_{0} /\omega \mu _{0} $ and $\textbf{k}=-\omega \hat{s}/c$. In both solutions, we may set $\left|\textbf{H}_{0} \right|=\left|\textbf{E}_{0} \right|/\eta _{0} =1$ and $\measuredangle E_{0} =\measuredangle H_{0} =0$ for the sake of simplicity. Now, by setting $\left(\hat{s},\hat{e},\hat{g}\right)=\left(\hat{x},\hat{y},\hat{z}\right)$ we get

\begin{eqnarray}
 {\textbf{E}\left(\textbf{r},t\right)=\hat{y}\eta _{0} \cos \left[\omega \left(t-\frac{1}{c} x\right)\right]}  \\
 {\textbf{H}\left(\textbf{r},t\right)=\hat{z}\cos \left[\omega \left(t-\frac{1}{c} x\right)\right]} \nonumber
\end{eqnarray}

\noindent as the right-handed solution. Similarly, we may take $\left(\hat{s},\hat{e},\hat{g}\right)=\left(\hat{x},\hat{z},\hat{y}\right)$ to get

\begin{eqnarray}
{\textbf{E}\left(\textbf{r},t\right)=\hat{z}\eta _{0} \cos \left[\omega \left(t+\frac{1}{c} x\right)\right]} \\
{\textbf{H}\left(\textbf{r},t\right)=\hat{y}\cos \left[\omega \left(t+\frac{1}{c} x\right)\right]} \nonumber
\end{eqnarray}

\noindent as the left-handed solution.

It is easy to verify by direct substitution in (A1), that (A5) and (A6) are indeed both correct solutions of Maxwell's equations. By inspection, it may be easily seen that (A5) represents a wave that propagates along $+x$ ($-x$) when time increases (decreases). At the same time, (A6) represents a wave that propagates along $+x$ ($-x$) when times decreases (increases). Hence these two solutions may be better coined as waves traveling forward and backward in time, corresponding to the trivial refractive indices $n_{1} =+1$ and $n_{2} =-1$, respectively, represented as

\begin{eqnarray}
{\textbf{E}_{1} \left(\textbf{r},t\right)=\hat{y}\eta _{0} \cos \left[\omega \left(t-\frac{n_{1} }{c} x\right)\right]} \\
{\textbf{H}_{1} \left(\textbf{r},t\right)=\hat{z}\cos \left[\omega \left(t-\frac{n_{1} }{c} x\right)\right]} \nonumber
\end{eqnarray}

\noindent and

\begin{eqnarray}
{\textbf{E}_{2} \left(\textbf{r},t\right)=\hat{z}\eta _{0} \cos \left[\omega \left(t-\frac{n_{2} }{c} x\right)\right]} \\
{\textbf{H}_{2} \left(\textbf{r},t\right)=\hat{y}\cos \left[\omega \left(t-\frac{n_{2} }{c} x\right)\right]} \nonumber
\end{eqnarray}

In conclusion, we have the following:

\noindent 1)  The apparently negative refractive index $-1$ of simple vacuum, corresponds to the waves traveling backward in time.

\noindent 2) Both solutions appear as right-handed if the triad is taken as $\left(\hat{k},\hat{e},\hat{g}\right)$ instead of $\left(\hat{s},\hat{e},\hat{g}\right)$.

\noindent 3) Since the unit vector along Poynting's vector lies along $\hat{p}=\hat{e}\times \hat{g}$, then $\hat{p}\cdot \hat{k}=1$, or $\textbf{P}\cdot \textbf{k}>0$. In other words, there is no such ``negative refraction''. Notice that $\textbf{P}\cdot \hat{s}>0$ may not hold for (A8).

\end{document}